# Deep-recessed β-Ga$_2$O$_3$ delta-doped field effect transistors with in situ epitaxial passivation

Chandan Joishi, Zhanbo Xia, John S. Jamison, Shahadat H. Sohel, Roberto C. Myers, Saurabh Lodha, and Siddharth Rajan

*Abstract*—We introduce a deep-recessed gate architecture in β-Ga$_2$O$_3$ delta-doped field effect transistors for improvement in DC-RF dispersion and breakdown properties. The device design incorporates an unintentionally doped β-Ga$_2$O$_3$ layer as the passivation dielectric. To fabricate the device, the deep-recess geometry was developed using BCl$_3$ plasma based etching at ~5 W RIE to ensure minimal plasma damage. Etch damage incurred with plasma etching was mitigated by annealing in vacuum at temperatures above 600 °C. A gate-connected field-plate edge termination was implemented for efficient field management. Negligible surface dispersion with lower knee-walkout at high $V_{DS}$, and better breakdown characteristics compared to their unpassivated counterparts were achieved. A three terminal off-state breakdown voltage of 315 V, corresponding to an average breakdown field of 2.3 MV/cm was measured. The device breakdown was limited by the field-plate/passivation edge and presents scope for further improvement. This demonstration of epitaxially passivated field effect transistors is a significant step for β-Ga$_2$O$_3$ technology since the structure simultaneously provides control of surface-related dispersion and excellent field management.

Manuscript received XX-XX-XXXX; revised XX-XX-XXXX. This work was supported in part by the National Science Foundation (NSF ECCS-1809682), Department of Defense- Defense Threat Reduction Agency (Grant HDTRA11710034), ONR EXEDE MURI program, and the OSU Institute for Materials Research Seed Program. The review of this article was arranged by Editor xxxxxx.

Chandan Joishi is with the Department of Electrical and Computer Engineering, The Ohio State University, Columbus, OH 43210 USA, and also with the Department of Electrical Engineering, Indian Institute of Technology Bombay, Mumbai, MH 400076, India. (email: cjnits@gmail.com)
Zhanbo Xia and Shahadat H. Sohel are with the Department of Electrical & Computer Engineering, The Ohio State University, Columbus, OH 43210, USA.
John S. Jamison is with the Department of Materials Science and Engineering, The Ohio State University, Columbus, OH 43210 USA.
Roberto C. Myers is with the Department of Electrical and Computer Engineering, The Ohio State University, Columbus, OH 43210 USA, the Department of Materials Science and Engineering, The Ohio State University, Columbus, OH 43210 USA and also with the Department of Physics, The Ohio State University, Columbus, Ohio 43210, USA.
Saurabh Lodha is with the Department of Electrical Engineering, Indian Institute of Technology Bombay, Mumbai, MH 400076, India.
Siddharth Rajan is with the Department of Electrical and Computer Engineering, The Ohio State University, Columbus, OH 43210 USA, and also with the Department of Materials Science and Engineering, The Ohio State University, Columbus, OH 43210 USA. (email: rajan.21@osu.edu)

## I. INTRODUCTION

CURRENTLY, there is extensive interest for exploring β-Ga$_2$O$_3$ (band gap ~ 4.6 eV) based devices for next generation high power switching [1], [2] and radio-frequency (RF) electronics [3], [4]. This is primarily driven by the availability of superior quality, large area monocrystalline β-Ga$_2$O$_3$ bulk substrates with different orientations from low-cost melt based methods [5]-[7], a feature distinctive to β-Ga$_2$O$_3$ amongst all the wide band gap materials [8]. Besides, the ultra-wide band gap enables a high breakdown field (~8 MV/cm) [9] which when combined with high saturation velocity (1-2x10$^7$ cm/s) [10], [11] and electron mobility (250-350 cm$^2$/Vs) [12], quantifies to one of the best figure of merits for power switching [13], [14] and amplification [15] amidst contemporary wide band gap materials. This widespread interest in β-Ga$_2$O$_3$ has prompted research groups to develop lateral and vertical transport devices in the form of Schottky barrier diodes [16]-[21] and field-effect transistors (FETs) [22]-[32] with excellent performance indices.

Of the various lateral β-Ga$_2$O$_3$ based FETs demonstrated in literature, delta-doped FETs [4], [24], [33] were shown to manifest high sheet charge density at improved mobility, high ON-current density and unity current gain cut-off frequency besides appreciable breakdown characteristics at low ON-resistance (R$_{ON}$). The efficacy of these lateral FETs is however, limited by the peak electric field at the drain-side edge of the gate, as in all lateral wide band gap transistor topologies [34]. The exposed semiconductor surface in the gate-drain access region is susceptible to this high field that lead to charge trapping in surface states, thereby resulting in DC-RF dispersion and dynamic R$_{ON}$ degradation [35]- [36]. Although the presence of buffer traps also contribute to the process, charge trapping in surface states is argued to be the primary source of this degradation [35], [38]. For the rest of the paper, we refer to this degradation as dispersion.

To realize delta-doped FETs with excellent performance, dispersion incurred by the presence of deep-level states in the buffer from semi-insulating Fe doped substrate was reduced by using thick buffer layers [33]. To mitigate the surface-related dispersion for these FETs, passivation of the semiconductor surface is essential. Although silicon dioxide (SiO$_2$) [26], [27], [37] and silicon nitride (SiN$_x$) [32] have been investigated as passivation dielectrics on β-Ga$_2$O$_3$ FETs, the semiconductor-dielectric interface is far from ideal as the dielectrics were deposited ex situ. On the other hand, epitaxial passivation, wherein the dielectric is grown in situ with the device epitaxial stack to function as the passivation layer could outperform the ex situ deposited dielectrics due to their improved interfacial properties. The phenomenon is further enhanced for a delta-doped FET with unintentionally doped (UID) β-$Ga_2O_3$ being used as the device cap as well as the epitaxial passivation layer since the design ensures the absence of a semiconductor-dielectric interface near regions of the device that experience the highest electric fields. Furthermore, the high dielectric permittivity of the β-$Ga_2O_3$ passivation layer can enable superior breakdown characteristics when combined with optimal field-plate designs for field management. Indeed, epitaxial passivation previously investigated on gallium nitride high electron mobility transistors (GaN HEMTs) using UID GaN as the passivation dielectric demonstrated excellent dispersion and breakdown properties [39], [40].

This paper focuses on development of epitaxial passivation for β-$Ga_2O_3$ delta-doped FETs. Such a novel architecture would require etching of UID β-$Ga_2O_3$ underneath the gate to form a deep-recess pattern that differentiates the passivation layer from the cap layer. The deep-recess was patterned using $BCl_3$ plasma based dry etching. Damage incurred by $BCl_3$ plasma during deep-recess was found to degrade two-terminal source-drain current characteristics. Hence, we investigated techniques to restore current density after deep-recess. With optimal plasma etch conditions alongside a damage recovery anneal step; the source-drain current was restored and field-plated β-$Ga_2O_3$ delta-doped FETs with epitaxial passivation that exhibited negligible dispersion and enhanced breakdown characteristics were developed.

## II. DEVICE GROWTH AND FABRICATION DETAILS

The epitaxial stack for the delta-doped FET was grown in an oxygen plasma assisted molecular beam epitaxy (PAMBE) chamber on (010) oriented Fe-doped β-$Ga_2O_3$ semi-insulating substrates commercially available from Tamura Corporation [41]. The growth was realized in a gallium-deficient regime with a growth rate ~3 nm/min. An $O_2$ plasma power of 300 W, chamber pressure of ~$1.5 \times 10^{-5}$ Torr, gallium beam equivalent pressure (BEP) of ~$8 \times 10^{-8}$ Torr, and substrate temperature of 700 °C were maintained during the growth. Before initiating the growth, the substrate was heated to 800 °C and a surface cleaning process using $O_2$ plasma was carried out for 10 minutes. The delta-doping profile was realized with Si as the dopant. To delta dope UID β-$Ga_2O_3$, the Si cell shutter was opened for 3 sec with cell temperature maintained at 900 °C without growth interruption. This specific combination of the shutter time and cell temperature was chosen to target charge density ~$10^{13}$ $cm^{-2}$. Prior to setting the Si cell to the desired temperature, the cell was heated at 1150°C for 15 minutes to remove the residual oxide that forms on top of the Si target in the cell.

The epitaxial structure (starting from the substrate)

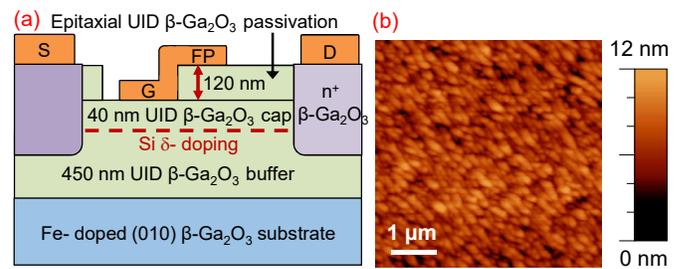

**Fig. 1.** (a) 2-D schematic of the fabricated deep-recessed delta-doped FET with UID β-$Ga_2O_3$ as the passivation layer grown in situ and a gate connected field-plate. (b) Atom force microscopy image of the as-grown β-$Ga_2O_3$ surface.

comprises a 450 nm UID β-$Ga_2O_3$ buffer, a Si delta-doped sheet, and 160 nm of UID β-$Ga_2O_3$ to function as the cap (40 nm) plus the passivation layer (120 nm). A 2-D schematic of the device epitaxial stack that also portrays the final deep-recessed design is shown in Fig. 1(a). The as-grown surface displayed a smooth morphology as measured using atom probe microcopy (AFM) with an rms roughness of 1.5 nm (Fig. 1(b)). To fabricate the device, source-drain ohmic contacts were first formed using a patterned regrowth process flow. In this process, a 500 nm thick sacrificial $SiO_2$ layer was deposited on the as-grown sample in a plasma enhanced chemical vapor deposition (PECVD) chamber at 250 °C. Source and drain regions were then lithographically defined on the $SiO_2$ layer. 470 nm of $SiO_2$ underneath the defined region was etched using $CF_4/O_2/Ar$ plasma based dry etch in a PlasmaTherm SLR 770 ICP-RIE system at 120 W RIE, 120 W ICP and a chamber pressure of 5 mTorr. The remaining 30 nm $SiO_2$ was conformally etched in a diluted buffer oxide etchant (BOE) to expose the β-$Ga_2O_3$ surface. Using $SiO_2$ as the hard mask, ~190 nm β-$Ga_2O_3$ underneath the pattern was etched to reach the delta-doping profile using $BCl_3$ plasma in ICP-RIE (20 sccm $BCl_3$, 30 W RIE, 200 W ICP, and 5 mTorr chamber pressure). Degenerately doped β-$Ga_2O_3$ with layer thickness ~50 nm and a targeted doping density of ~$10^{20}$ $cm^{-3}$ was then regrown in these recessed regions using PAMBE. After growth, β-$Ga_2O_3$ that rests on top of the sacrificial oxide was lifted off by etching $SiO_2$ in BOE. Following lift-off, $n^+$-$Ga_2O_3$ would exist only on the patterned source-drain contact regions. A Ti/Au (30nm/130nm) metal stack was then evaporated on the regrown contact regions, followed by $N_2$ anneal at 470 °C for 1 minute to form the ohmic metal contact. Mesa isolation was performed using $BCl_3$ plasma. To form the deep-recess pattern, UID β-$Ga_2O_3$ underneath the gate was etched using $BCl_3$ plasma to segregate the passivation layer from the cap layer. The process incurred substantial plasma damage causing degradation of the two-terminal source-drain current. This etch induced damage was subsequently recovered by using a damage recovery anneal step which is discussed in detail in section III. Thereafter, formations of gate and field-plate terminals constituted the final fabrication step. The inherent nature of the deep-recess pattern allowed for a controlled misalignment of the gate mask (towards the drain) on top of the recessed area to simultaneously define the Schottky gate and the field-plate contact in the same lithography process.

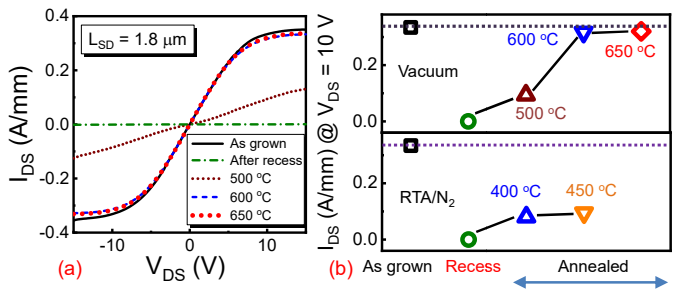

**Fig. 2.** (a) Two terminal drain-source characteristics ($I_{DS}$) of the as grown, the deep-recessed and the vacuum annealed (after recess) samples. The current recovery is plotted for $V_{DS}$ = 10 V in (b) for the vacuum annealed and RTA annealed (in $N_2$ ambient) samples. Near complete recovery of initial $I_{DS}$ was achieved in vacuum.

With a gate mask misalignment of 0.4 μm that corresponds to the field-plate length ($L_{FP}$), e-beam deposited Ni/Au (30 nm/100 nm) was used to form the gate plus the field-plate metal contact.

## III. ETCH DAMAGE CONTROL

To define the passivation layer, 120 nm of UID β-$Ga_2O_3$ beneath the gate region was etched using $BCl_3$ plasma in an ICP-RIE chamber. A chamber pressure of 5 mTorr, 20 sccm $BCl_3$, 30 W RIE, and 200 W ICP (etch rate = 32 nm/min) were initially adopted as the etching parameters. A drastic reduction in the two terminal source-drain current density (from ~0.3 A/mm to ~μA/mm) was observed after the recess. The degradation was attributed to damage incurred by $BCl_3$ plasma during the etch, also reported for ($\bar{2}$01) β-$Ga_2O_3$ [42] and other semiconductor technologies [43]-[45]. To reduce the damage due to $BCl_3$ plasma, a minimal plasma intensity of ~5 W RIE (40 V DC bias) with no ICP power was used at a chamber pressure of 5 mTorr, and $BCl_3$ flow of 20 sccm. The recipe resulted in an etch rate of ~2.7 nm/min. Although low plasma damage was anticipated for the recipe, current degradation was still observed after the recess. Further investigation revealed that the degradation was independent of plasma power, chamber pressure and $BCl_3$ flow rate.

For ($\bar{2}$01) β-$Ga_2O_3$ Schottky barrier diodes, Yang et al. [42] had observed that annealing at 400-450 °C for 10 minutes in an Ar ambient was effective towards passivating the damage incurred by the $BCl_3$/Ar plasma. Therefore, to revive the source-drain current after recess, the etched devices were annealed at 400 °C for 10 minutes in $N_2$ ambient in a rapid thermal anneal (RTA) chamber. However, only 20-30% of the initial current density could be recovered after anneal. Increasing the annealing temperature ($T_{anneal}$) to 450 °C showed a negligible increase in drain current compared to 400 °C. Further increase in $T_{anneal}$ to 500 °C was seen to degrade the Ti/Au ohmic contact. Similar results were also observed for anneal in Ar ambient. Since annealing at atmospheric pressure in $N_2$ or Ar ambient did not enable recovery of current density, we attempted vacuum annealing at a pressure of ~3.5x$10^{-7}$ Torr for a longer duration (60 minutes).

A typical two-terminal source-drain current profile for a 1.8 μm source-drain separation for the sample as-grown, after recess and after annealing is shown in Fig. 2(a). Annealing in vacuum was seen to enable higher $T_{anneal}$ without degradation of the ohmic contact. More than 90% of the initial current density could be recovered at 600 °C. Beyond 600 °C, a negligible increase in the current density was observed, i.e. the revived current was seen to nearly saturate after $T_{anneal}$ > 600 °C. The vacuum anneal was also experimented on the recess etches carried out with high RIE (30 W) and ICP (200 W) powers. However, only 50% of the total current density could be recovered. This implies that the current recovery index is a function of the plasma power, the anneal temperature and pressure. Further optimization is required for complete recovery of the depleted current. The current recovery for both vacuum and $N_2$ anneal is plotted for a typical source-drain voltage ($V_{DS}$) of 10 V in Fig. 2(b).

Hall measurements were carried out at room temperature before the deep-recess to estimate the sheet charge density. A Hall mobility of 73 $cm^2$/Vs at a sheet charge density of 1.2x$10^{13}$ $cm^{-2}$ was measured. Degradation in source-drain current after the recess (before anneal) was observed to be in agreement with reduction in the Hall sheet charge density from 1.2x$10^{13}$ $cm^{-2}$ to ~$10^{10}$ $cm^{-2}$. To qualitatively understand this phenomenon of charge degradation with recess, photoluminescence (PL) measurements were carried out with an excitation laser wavelength of 232 nm. The PL spectra of a bare Fe doped β-$Ga_2O_3$ substrate, the as-grown delta-doped epitaxial stack, the stack after recess and after anneal are plotted in Fig 3(a). As seen from the figure, intensity of PL transitions that exist for the as-grown sample diminishes after deep-recess. The spectrum however, is seen to reappear for the 'after anneal' sample. To zero down to the wavelengths involved, the individual spectrum for the 'as-grown' and the 'after anneal' samples are plotted in Fig. 3(b) and (c) with Gaussian distribution to fit the experimental data. The Gaussian fits for the "as-grown" spectrum de-convolved the emission band into three dominant peaks at 370 nm (3.36 eV), 410 nm (3 eV), and 480 nm (2.6 eV). These transition peaks are attributed to radiative recombination of electrons from the conduction band edge with holes in the acceptor bands (gallium vacancies) that give rise to UV-blue emissions as reported in literature [46]. These peaks were extensive for the as-grown sample but diminished for the deep-recessed sample. The observation is consistent with electrons being trapped at deep level states that result from damage due to $BCl_3$ plasma for the 'after-recessed' sample. The depletion in free electrons in turn, results in a decrease in PL transitions thereby decreasing the PL intensity. The theory is further supported by a decrease in the Hall sheet charge density after recess. The annealed sample showed almost identical peak positions and intensity in the PL spectrum, suggesting that annealing at high temperatures was effective towards mitigating the deep level states and electron trapping in the epitaxial layers caused by $BCl_3$ plasma. The Hall sheet charge density also recovered to $\sim 10^{13}$ cm$^{-2}$ after annealing. Although a detailed investigation is required to quantitatively confirm the hypothesis, similar observations for plasma damage during etch and recovery after anneal for other semiconductor technologies do exist in literature [45].

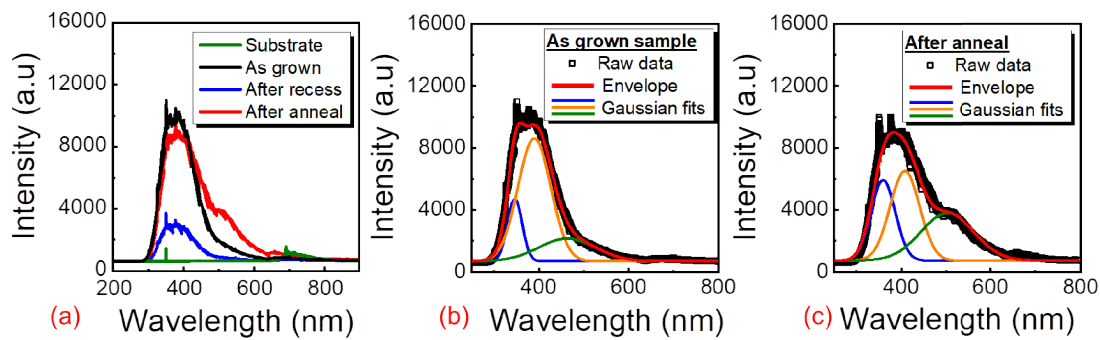

**Fig. 3.** (a) Room temperature PL spectra of an Fe doped (010) β-$Ga_2O_3$ template, the as grown epitaxial stack, the stack after deep recess and after anneal. Gaussian peak fits of PL spectra are shown for (b) the as grown sample and (c) the sample after anneal. Three dominant transitions at ~370 nm, ~410 nm, and ~480 nm are seen. These transitions are suppressed in the PL spectra of the recessed sample as seen in (a) while they reoccur after anneal.

## IV. RESULTS AND DISCUSSIONS

With formation of gate and field-plate terminals after the recovery of source-drain current, the electrical characteristics of the device were measured using a Keysight 1500A semiconductor device parameter analyzer. The three terminal transfer characteristics of the device ($I_{DS}$-$V_{GS}$) with $L_{SD}$ (separation between source-drain regrowth edges) = 5.4 µm, $L_G$ (gate length) = 0.65 µm, and $L_{GD}$ (gate to drain regrowth edge separation) = 2.7 µm for a device width of 75 µm is shown in Fig. 4(a). The dimensions were confirmed from scanning electron microscope (SEM) images. A high $I_{ON}/I_{OFF}$ ratio of $10^7$, pinch-off voltage of -10 V, and a low off-state leakage current of $3 \times 10^{-8}$ A/mm were measured for the device. The output curves ($I_{DS}$-$V_{DS}$) of the device is shown in Fig. 4(b). A saturated drain current density of 180 mA/mm was measured $V_{GS}$ = 2 V and $V_{DS}$ = 15 V. This current density was found to increase with decrease in $L_{SD}$.

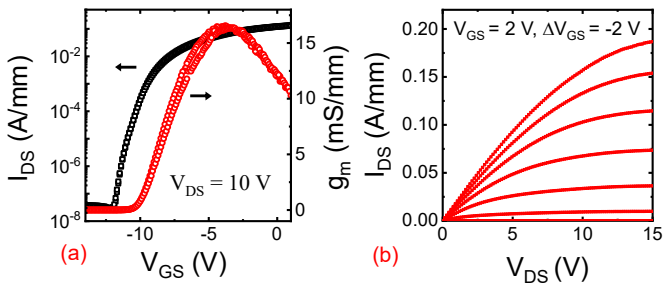

Fig. 4. (a) $I_{DS}$-$V_{GS}$ and (b) $I_{DS}$-$V_{DS}$ characteristics of the deep-recessed FET. The device exhibits a pinch-off voltage of -10 V, an $I_{ON}/I_{OFF}$ ratio of $10^7$, and a maximum $I_{ON}$ of ~180 mA/mm.

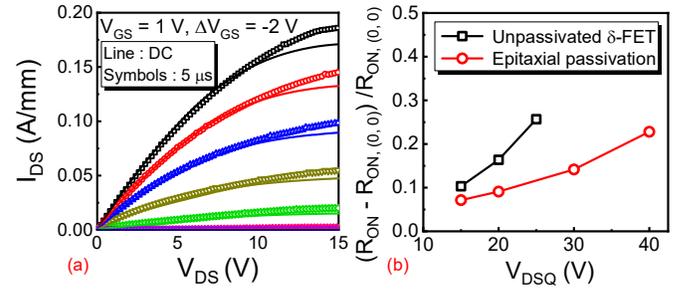

Fig. 5. (a) DC and pulsed $I_{DS}$-$V_{DS}$ curves at OFF-state bias ($V_{GSQ}$, $V_{DSQ}$) = (-15 V, 15 V). (b) Normalized change in $R_{ON}$ vs OFF-state $V_{DSQ}$ is compared to an unpassivated delta-doped FET.

To investigate for dispersion, pulsed I-V measurements were carried out in a Keithley 4200-SCS parameter analyzer with an on-state pulse width of 5 µs and 0.1% duty cycle. The off-state quiescent bias point was set at $V_{GSQ}$ = -15 V (below pinch-off) and $V_{DSQ}$ = 15 V (at saturation), where Q in the subscript differentiates a pulsed voltage from a DC voltage. Negligible dispersion was observed for the devices biased at pulsed IV conditions as shown in Fig. 5(a). As expected, the pulsed IV drain current at saturation ($I_{DSQ, sat}$) was higher than the corresponding DC values due to the absence of self-heating. The dispersion characteristics were better compared to the unpassivated delta-doped FETs reported in [33]. This is primarily due to the fact that the exposed/unpassivated surface in the gate-drain access region for the deep-recessed FET, i.e. the top facet of the passivation layer, is far-off from the delta-doped channel. The surface states in this facet are not prone to high electric fields from the gate edge. The large separation as such, is effective towards suppressing the trapping effects on the delta-doped channel from the surface states. The phenomenon is further enhanced by the absence of a definite β-$Ga_2O_3$/dielectric interface on the gate-drain plane that is exposed to high fields, which is precisely the situation for the surface states of the unpassivated delta-doped FETs.

To estimate the degree of knee walkout/ on-resistance ($R_{ON}$) degradation, difference in $R_{ON}$ ($\Delta R_{ON}$) for different ($V_{GSQ}$, $V_{DSQ}$) off-state bias points normalized to $R_{ON, (0, 0)}$ ($R_{ON}$ extracted for ($V_{GSQ}$, $V_{DSQ}$) = (0, 0)) for the deep-recessed epitaxially passivated FET is compared to an unpassivated delta-doped FET in Fig. 5(b). $R_{ON}$ was extracted from the linear region of the $I_{DS}$-$V_{DS}$ curve with $V_{GS}$ at 0 V while $\Delta R_{ON}$ was calculated using the equation,

$$\Delta R_{ON} = R_{ON} - R_{ON, (0, 0)} \quad (1)$$

The pulsed voltage compliance of the measuring instrument limited the extraction of $R_{ON}$ for the deep-recessed FET to $V_{DSQ}$ = 40 V. For the unpassivated FET, the measurement was done till $V_{DSQ}$ = 25 V, since the device presented appreciable gate leakage beyond 25 V.Compared to the steep increase in $\Delta R_{ON}$ for the unpassivated sample, the percentage change in $R_{ON}$ is lesser for the deep-recessed sample, i.e. the devices are more robust towards $R_{ON}$ degradation. The increase in $R_{ON}$ at high $V_{DSQ}$ for the deep-recessed FET is possibly due to the presence of bulk/buffer traps in the material that are activated at high fields and require further analysis.

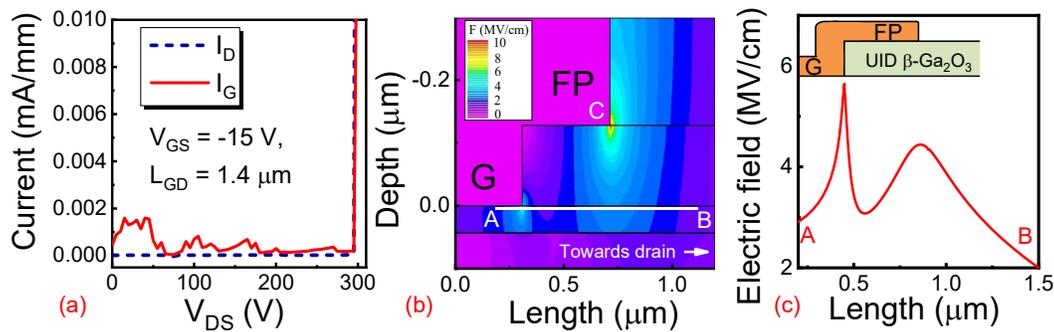

**Fig. 6.** The three terminal OFF-state breakdown characteristics for the field-plated deep-recessed FET. (b) 2-D contour plot showing the electric field profile for the deep-recessed FET at breakdown. (c) Electric field profile beneath the gate extending to the drain (along AB).

To examine the breakdown properties of the deep-recessed FETs, the three-terminal off-state voltage breakdown was measured using a Keysight B1505A power device analyzer. The measurement was initiated after immersing the device under test in few drops of Fluorinert liquid to prevent air breakdown. For $L_{GD}$ =1.4 µm with the gate biased at -15 V, the device broke down catastrophically at $V_{DS}$ = 300 V (Fig. 6(a)). This translates to an average breakdown field of 2.3 MV/cm. On the other hand, an unpassivated FET with similar charge density (~$10^{13}$ cm$^{-2}$) and architecture gave a Schottky limited breakdown of 164 V for the same $L_{GD}$ [4]. The deep-recessed FET therefore, exhibits superior breakdown characteristics compared to its unpassivated counterpart.

2-D TCAD simulations were performed using SILVACO Atlas to estimate electric fields within the device at the onset of breakdown. A 2-D contour plot of the simulated electric field profile at $V_{DS}$ = 300 V and $V_{GS}$ = -15 V is plotted in Fig. 6(b). As shown in the plot, the field peaks at the field-plate/passivation edge (point C in the graph). The validation of field-plate action for the deep-recess design is shown in Fig. 6(c) along a horizontal line 'AB' from gate to drain. The total potential applied to the drain is normalized between the gate and the region below the field-plate edge. Since the peak field hotspot is not located at the gate edge (drain side), the device breakdown is limited by the field-plate/passivation layer edge and can be increased further through additional engineering of the passivation layer design and electrostatic fields within the device.

## V. CONCLUSIONS

In summary, we have demonstrated a deep-recessed delta-doped FET with UID β-Ga$_2$O$_3$ as the passivation dielectric grown in-situ on the device epitaxial stack. Current depletion observed during the deep recess etch of β-Ga$_2$O$_3$ by BCl$_3$ plasma, attributed to plasma damage that contribute to the emergence of deep level states in the material after recess, was mitigated through the use of a damage recovery anneal step. The FETs displayed negligible dispersion and better breakdown features compared to the unpassivated FETs, with scope for further improvement. The device architecture developed here is indeed promising to explore the potential of β-Ga$_2$O$_3$ for next generation power, RF and mm-Wave device applications.

REFERENCES

[1] M. H. Wong *et al.*, "Gallium Oxide Field Effect Transistors — Establishing New Frontiers of Power Switching and Radiation-Hard Electronics." *International Journal of High Speed Electronics and Systems,* vol. 28, no. 01n02, pp. 1940002, 2019. DOI: 10.1142/S0129156419400020

[2] J. Y. Tsao *et al.*, "Ultrawide-bandgap semiconductors: Research opportunities and challenges", *Adv. Electron. Mater.*, vol. 4, no. 1, pp. 1600501, 2018, DOI: 10.1002/aelm.201600501

[3] A. J. Green *et al.*, "β-Ga$_2$O$_3$ MOSFETs for Radio Frequency Operation," in *IEEE Electron Device Letters*, vol. 38, no. 6, pp. 790-793, June 2017. DOI: 10.1109/LED.2017.2694805

[4] Z. Xia *et al.*, " β-Ga$_2$O$_3$ Delta-Doped Field-Effect Transistors With Current Gain Cutoff Frequency of 27 GHz," in *IEEE Electron Device Letters*, vol. 40, no. 7, pp. 1052-1055, July 2019.DOI: 10.1109/LED.2019.2920366

[5] T. Oishi, *et al.*, "High-mobility β-Ga$_2$O$_3$ ($\bar{2}$01) single crystals grown by edge-defined film-fed growth method and their Schottky barrier diodes with Ni contact," *Appl. Phys. Exp.,* vol. 8, no. 3, p. 031101, 2015, DOI: 10.7567/APEX.8.031101

[6] Z. Galazka *et al.*, "Czochralski growth and characterization of β-Ga$_2$O$_3$ single crystals," *Crystal Res. Technol.*, vol. 45, no. 12, pp. 1229–1236, 2010, DOI: 10.1002/crat.201000341.

[7] E. G. Víllora *et al.*, "Large-size β-Ga$_2$O$_3$ single crystals and wafers," *J. Crystal Growth*, vol. 270, no. 3–4, pp. 420–426, Oct. 2004. DOI: 10.1016/j. jcrysgro.2004.06.027.

[8] S. J. Pearton *et al.*, "A review of Ga$_2$O$_3$ materials, processing, and devices." *Applied Physics Reviews, vol.* 5, no. 1 pp. 011301, 2018. DOI: 10.1063/1.5006941

[9] M. Higashiwaki *et al.*, "Gallium oxide (Ga$_2$O$_3$) metal-semiconductor field-effect transistors on single-crystal β-Ga$_2$O$_3$ (010) substrates," *Applied Physics Letters*, vol. 100, pp. 013504, 2012. DOI: 10.1063/1.3674287.

[10] K. Ghosh *et al.*, "Ab initio velocity-field curves in monoclinic β-Ga$_2$O$_3$," *J. Appl. Phys.*, vol. 122, no. 3, pp. 035702, 2017, DOI: 10.1063/1.4986174.

[11] Y. Zhang *et al.*, "Evaluation of Low-Temperature Saturation Velocity in β-(Al$_x$Ga$_{1-x}$)$_2$O$_3$/Ga$_2$O$_3$ Modulation-Doped Field-Effect Transistors," in *IEEE Transactions on Electron Devices*, vol. 66, no. 3, pp. 1574-1578, March 2019. DOI: 10.1109/TED.2018.2889573

[12] N. Ma *et al.*, "Intrinsic electron mobility limits in β-Ga$_2$O$_3$," *Appl. Phys. Lett.*, vol. 109, no. 21, pp. 212101, Nov. 2016. DOI: 10.1063/1.4968550

[13] B. J. Baliga, "Power semiconductor device figure of merit for high-frequency applications," in *IEEE Electron Device Letters*, vol. 10, no. 10, pp. 455-457, Oct. 1989. DOI: 10.1109/55.43098

[14] A. Q. Huang, "New unipolar switching power device figures of merit," *IEEE Electron Device Letters*, vol. 25, no. 5, pp. 298–301, 2004. DOI:  10.1109/LED.2004.826533

[15] E. Johnson, "Physical limitations on frequency and power parameters of transistors," *1958 IRE International Convention Record*, New York, NY,

USA, 1965, pp. 27-34. DOI: 10.1109/IRECON.1965.1147520

[16] K. Konishi *et al.*, "1-kV vertical Ga$_2$O$_3$ field-plated Schottky barrier diodes," *Appl. Phys. Lett.*, vol. 110, no. 10, p. 103506, 2017, DOI: 10.1063/1.4977857

[17] C. Joishi *et al.*, "Low-pressure CVD-grown β-Ga$_2$O$_3$ bevel-field-plated Schottky barrier diodes," in *Appl. Phys. Exp,* vol. 11, no. 3, pp. 031101, Feb 2018. DOI: 10.7567/APEX.11.031101

[18] K. Sasaki *et al.*, "First Demonstration of Ga$_2$O$_3$ Trench MOS-Type Schottky Barrier Diodes," in *IEEE Electron Device Letters*, vol. 38, no. 6, pp. 783-785, June 2017. DOI: 10.1109/LED.2017.2696986

[19] Z. Hu *et al.*, "Field-Plated Lateral β-Ga$_2$O$_3$ Schottky Barrier Diode With High Reverse Blocking Voltage of More Than 3 kV and High DC Power Figure-of-Merit of 500 MW/cm$^2$," in *IEEE Electron Device Letters*, vol. 39, no. 10, pp. 1564-1567, Oct. 2018. DOI: 10.1109/LED.2018.2868444

[20] W. Li *et al.*, "Field-Plated Ga$_2$O$_3$ Trench Schottky Barrier Diodes With a BV$^2$/ R$_{on,sp}$ of up to 0.95 GW/cm$^2$," in *IEEE Electron Device Letters*, vol. 41, no. 1, pp. 107-110, Jan. 2020. DOI: 10.1109/LED.2019.2953559

[21] C. Lin *et al.*, "Vertical Ga$_2$O$_3$ Schottky Barrier Diodes With Guard Ring Formed by Nitrogen-Ion Implantation," in *IEEE Electron Device Letters*, vol. 40, no. 9, pp. 1487-1490, Sept. 2019. DOI: 10.1109/LED.2019.2927790

[22] M. H. Wong *et al.*, "Enhancement-Mode β-Ga$_2$O$_3$ Current Aperture Vertical MOSFETs With N-Ion-Implanted Blocker," in *IEEE Electron Device Letters*, vol. 41, no. 2, pp. 296-299, Feb. 2020. DOI: 10.1109/LED.2019.2962657

[23] A. J. Green *et al.*, "3.8-MV/cm breakdown strength of MOVPE-grown Sn-doped β-Ga$_2$O$_3$ MOSFETs," *IEEE Electron DeviceLett.*, vol. 37, no. 7, pp. 902–905, Jul. 2016, DOI: 10.1109/LED.2016.2568139.

[24] Z. Xia *et al.*, "Delta doped β-Ga$_2$O$_3$ Field Effect Transistors with Regrown Ohmic Contacts," in *IEEE Electron Device Letters*, vol. PP, no. 99, pp. 1-1. DOI: 10.1109/LED.2018.2805785

[25] M. H. Wong *et al.*, "Field-plated Ga$_2$O$_3$ MOSFETs with a breakdown voltage of over 750 V," *IEEE Electron Device Lett.,* vol. 37, no. 2, pp. 212–215, Feb. 2016, DOI: 10.1109/LED.2015.2512279.

[26] K. Zeng *et al.*, "1.85 kV Breakdown Voltage in Lateral Field-Plated Ga$_2$O$_3$ MOSFETs," in *IEEE Electron Device Letters*, vol. 39, no. 9, pp. 1385-1388, Sept. 2018. DOI: 10.1109/LED.2018.2859049

[27] J. K. Mun *et al.*, "2.32 kV Breakdown Voltage Lateral β-Ga$_2$O$_3$ MOSFETs with Source-Connected Field Plate", *ECS Journal of Solid State Science and Technology*, vol. 8, no. 7, pp. Q3079-Q3082. DOI: 10.1149/2.0151907jss

[28] Y. Zhang *et al.*, "Demonstration of high mobility and quantum transport in modulation-doped β-(Al$_x$Ga$_{1-x}$) $_2$O$_3$/Ga$_2$O$_3$ heterostructures." in *Appl. Phys. Lett.,*vol 112, no. 17, pp. 173502, 2018. DOI: 10.1063/1.5025704

[29] S. Krishnamoorthy *et al.*, "Modulation-doped β-(Al$_{0.2}$Ga$_{0.8}$) $_2$O$_3$/Ga$_2$O$_3$ field-effect transistor," in *Appl. Phys. Lett.*, vol. 111, no. 2, pp. 023502, 2017. DOI: 10.1063/1.4993569

[30] Y. Zhang *et al.*, "Demonstration of β-(Al$_x$Ga$_{1-x}$)$_2$O$_3$/Ga$_2$O$_3$ double heterostructure field effect transistors." in *Appl. Phys. Lett.,* vol. 112, no. 23, pp. 233503, 2018. DOI: 10.1063/1.5037095

[31] Z. Hu *et al.*, "Enhancement-Mode Ga$_2$O$_3$ Vertical Transistors with Breakdown Voltage >1 kV," in *IEEE Electron Device Letters*, vol. 39, no. 6, pp. 869-872, June 2018. DOI: 10.1109/LED.2018.2830184

[32] C. Joishi *et al.*, "Breakdown Characteristics of β-(Al$_{0.22}$Ga$_{0.78}$)$_2$O$_3$/Ga$_2$O$_3$ Field-Plated Modulation-Doped Field-Effect Transistors," in *IEEE Electron Device Letters*, vol. 40, no. 8, pp. 1241-1244, Aug. 2019. DOI: 10.1109/LED.2019.2921116

[33] C Joishi *et al.*, "Effect of buffer iron doping on delta-doped β-Ga$_2$O$_3$ metal semiconductor field effect transistors," in *Appl. Phys. Lett.*, vol. 113, no. 12, pp. 123501, 2018. DOI: 10.1063/1.5039502

[34] Y. Dora, "Understanding material and process limits for high breakdown voltage AlGaN/GaN HEMTs," Ph.D. dissertation, University of California, Santa Barbara, 2006

[35] R. Vetury *et al.*, "The impact of surface states on the DC and RF characteristics of AlGaN/GaN HFETs," IEEE *Transactions on Electron Devices*, vol. 48, no. 3, pp. 560–566, March 2001. DOI: 10.1109/16.906451

[36] W. Saito *et al.*, "Suppression of dynamic on-resistance increase and gate charge measurements in high-voltage GaN-HEMTs with optimized field-plate structure," *IEEE Trans. Electron Devices*, vol. 54, no. 8, pp. 1825- 1830, August 2007. DOI: 10.1109/TED.2007.901150

[37] D. Biswas *et al.*, "Enhanced n-type β-Ga$_2$O$_3$ ($\bar{2}$01) gate stack performance using Al$_2$O$_3$/SiO$_2$ bi-layer dielectric," *Applied Physics Letters*, vol. 114, no. 21, pp. 212106, May 2019. DOI: 10.1063/1.5089627

[38] D. Jin *et al.*, "Mechanisms responsible for dynamic ON-resistance in GaN high-voltage HEMTs," *2012 24th International Symposium on Power Semiconductor Devices and ICs*, Bruges, 2012, pp. 333-336. DOI: 10.1109/ISPSD.2012.6229089

[39] S. Wienecke *et al.*, "N-Polar GaN Cap MISHEMT With Record Power Density Exceeding 6.5 W/mm at 94 GHz," in *IEEE Electron Device Letters*, vol. 38, no. 3, pp. 359-362, March 2017. DOI: 10.1109/LED.2017.2653192

[40] B. Romanczyk, "Demonstration of Record-High mm-Wave Power Performance using N-Polar Gallium Nitride HEMTs," *Journal of the Microelectronic Engineering Conference*. vol. 25, no. 1, article 24, 2019. Available: https://scholarworks.rit.edu/ritamec/vol25/iss1/24

[41] [Online]. Available: http://www.tamurass.co.jp/en/products/gao/index.html

[42] J. Yang *et al.*, "Annealing of dry etch damage in metallized and bare (-201) Ga$_2$O$_3$." *Journal of Vacuum Science & Technology B, Nanotechnology and Microelectronics: Materials, Processing, Measurement, and Phenomena*. vol. 35, no. 5, pp. 051201, 2017. DOI: 10.1116/1.4986300

[43] M. E. Lin *et al.*, "Reactive ion etching of GaN using BCl$_3$." *Applied Physics Letters,* vol. 64, no. 7, pp. 887-888, 1994. DOI: 10.1063/1.110985

[44] R. J. Shul *et al.*, "Plasma-induced damage of GaAs pn-junction diodes using electron cyclotron resonance generated Cl$_2$/Ar, BCl$_3$/Ar, Cl$_2$/BCl$_3$/Ar, and SiCl$_4$/Ar plasmas." *Journal of Vacuum Science & Technology B: Microelectronics and Nanometer Structures Processing, Measurement, and Phenomena*. vol. 13, no. 1, pp. 27-33, 1995. DOI: 10.1116/1.587980

[45] J. Lee *et al.*, "Inductively coupled Ar plasma damage in AlGaAs," *Journal of The Electrochemical Society*, vol. 144, no. 9, pp. L245–L247, 1997. DOI: 10.1149/1.1837932

[46] W. Mi *et al.*, "Ultraviolet–green photoluminescence of β-Ga$_2$O$_3$ films deposited on MgAl$_6$O$_{10}$ (100) substrate," *Optical Materials*, vol. 35, no. 12, pp. 2624–2628, 2013. DOI: 10.1016/j.optmat.2013.07.030